# Electric-field control of magnetism in iron oxide nanoparticle / BaTiO$_3$ film composites

L.-M. Wang,[1,2,3] O. Petracic,[1,*] J. Schubert[4] and Th. Brückel[1]

[1]*Jülich Centre for Neutron Science JCNS and Peter Grünberg Institut PGI, JARA-FIT, Forschungszentrum Jülich GmbH, 52425 Jülich, Germany.*
[2]*Institute of High Energy Physics, Chinese Academy of Sciences, Beijing 100049, China*
[3]*Dongguan Neutron Science Center, Dongguan 523803, China*
[4]*Peter Grünberg Institute (PGI9-IT), JARA-FIT, Forschungszentrum Jülich GmbH, 52425 Jülich, Germany*

[*]*Corresponding author, email: o.petracic@fz-juelich.de*

**Abstract**

We study composites of monodisperse ferrimagnetic nanoparticles (NPs) embedded into ferroelectric barium titanate (BTO) films. The BTO films were prepared by pulsed laser deposition. The composite consists of a stack of two BTO films sandwiching one monolayer of iron oxide NPs. We observe a magnetoelectric coupling due to strain and interface charge co-mediation between the BTO and the NPs. This is demonstrated by measurements of the magnetization as function of DC and AC electric fields.

**Introduction**

Magnetoelectric (ME) materials allow the manipulation of the magnetization $M$ by an electric field $E$ or of the polarization by a magnetic field. ME materials have stimulated great research interest due to their potential applications in spintronics or for multifunctional devices [1-4].

Recently, a strain mediated ME coupling (MEC) effect was evidenced in various systems consisting of a FM thin film on top of a FE substrate, e.g. a $Fe_3O_4$ thin film on top of a Barium Titanate (BTO) substrate [5], or a $Ni_{77}Fe_{23}$ thin film on top of a $Pb(Mg_{1/3}Nb_{2/3})O_3$-$PbTiO_3$ (PMN-PT) substrate [6]. Considerable studies have also addressed systems composed of FM and FE bilayers. E.g. $La_{0.7}Sr_{0.3}MnO_3$ / $BaTiO_3$ [7], $La_{1.2}Sr_{1.8}Mn_2O_7$ /$PbZr_{0.3}Ti_{0.7}O_3$ [8], or $Ni_{0.5}Zn_{0.5}Fe_2O_4$ / $BaTiO_3$ [9]. Here the MEC effect can be ascribed to a strain or/and interface charge mediated coupling mechanism depending on the thickness of the FM layer [10, 11].

On the other hand, nanoparticles (NPs) can be regarded as building blocks for artificial super-structures [12, 13]. Such systems are particularly interesting because of their prospective applications in multifunctional materials. FM-FE or FiM-FE composites based on magnetic NPs are promising novel candidates for artificial nanoscale multiferroic devices [14, 15]. However, studies of ME systems consisting of self-assembled NPs are very limited and consequently several open questions remain. Therefore, we investigated the MEC effect in a composite system of self-assembled iron oxide NPs embedded in BTO films.

**Material and methods**

$5 \times 5 \times 0.5$ mm$^3$ (001) Niobium doped Strontium Titanate (Nb-doped STO) substrates (MaTeck) were used in this study. BTO films were deposited onto Nb-doped STO substrates by pulsed laser deposition (PLD) with 5 J/cm$^2$ energy density per pulse at 10 Hz repetition rate with a thickness of 500 nm. On top of the BTO film a layer of Iron oxide NPs coated with an oleic acid shell initially dispersed in toluene (Ocean NanoTech) was applied. The particles have a mean diameter of 20 nm and a size distribution width of 7%. The composition is hereby a mixture of γ-$Fe_2O_3$ and $FeO_x$ [16, 17]. 2.5 $\mu L$ diluted iron oxide NPs were taken by a transferpettor, dropped onto the BTO film and then spin-coated with 3000 rpm for 0.5 min. In order to remove the oleic acid shell around the NPs, the samples were then treated in an oxygen plasma asher (PNA Telpa 300, CNST) with an oxygen pressure of 0.3 mbar, a gas flow of 200 ml min$^{-1}$ and a power of 3000 W for 3 mins.

Subsequently a second 500 nm BTO film was deposited embedding the NPs. Finally, a 25 nm Au thin film was deposited via thermal evaporation serving as a top electrode, while the conductive Nb-doped STO substrate constitutes the bottom electrode for the application of an electric field across the BTO/NPs/BTO stack.

The crystalline quality of the as-prepared BTO films on Nb-doped STO substrates was characterized using an X-ray diffractometer (D8 advanced, Bruker AXS). The micromorphology of the NPs was studied using a scanning electron microscope (SEM, SU8000 from Hitachi). Magnetometry measurements were performed using a superconducting quantum interference device (SQUID) magnetometer (MPMS XL from Quantum Design). The magnetoelectric AC susceptibility (MEACS) was measured at the same SQUID magnetometer with an additional upgrade for the measurement of the ME coupling [18].

## Results

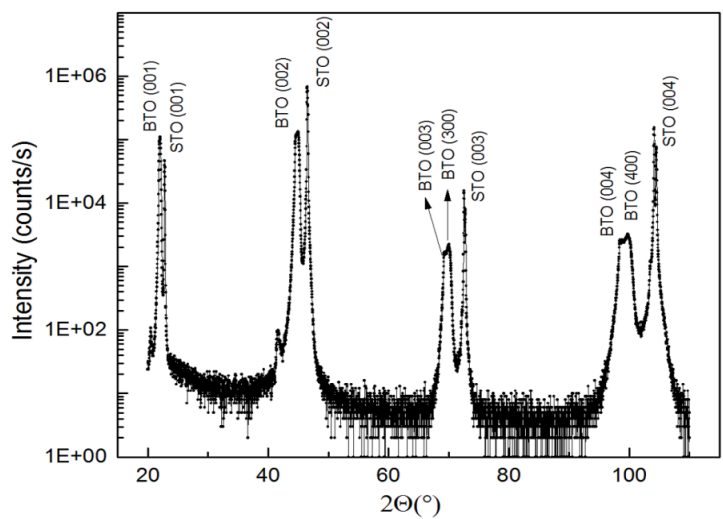

Figure 1. XRD pattern of the as-prepared BTO film on a Nb-doped STO substrate.

The X-ray pattern of the first as-prepared BTO film on top of the Nb-doped STO substrate is shown in Fig.1. The peaks are indexed according to Nb-doped STO and BTO lattice parameters. The result indicates that a ferroelectric BTO film is grown on the Nb-doped STO substrates. Split peaks are observable, i.e. the (004) and (400) reflections around 98°. They correspond to the tetragonal lattice constants c = 4.064 Å and a = 4.026 Å, respectively. These split peaks reveal that ferroelectric BTO films were prepared in the desired phase [19, 20].

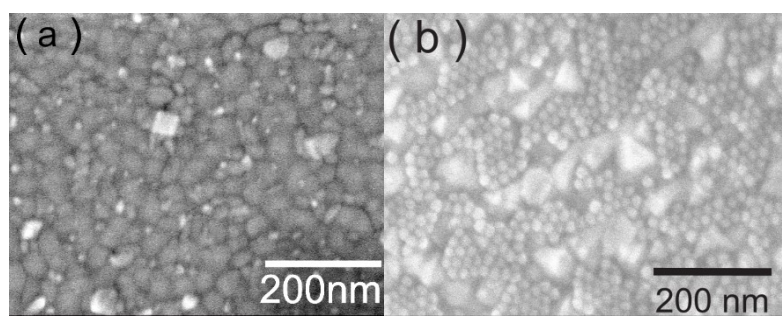

Figure 2. The morphology of (a) the as-prepared BTO film grown on a Nb-doped STO substrate and (b) iron oxide NPs self-assembled on top of the BTO film.

The topography of the as-prepared BTO films was studied using SEM as shown in Fig. 2(a). The NPs were then self-assembled on top of the BTO film using spin-coating. Fig. 2(b) shows the topography of the self-assembled NPs on top of the BTO film after oxygen plasma treatment. The triangular shaped spots are BTO islands. The iron oxide NPs assemble as a sub-monolayer in the valleys around the BTO islands.

To embed the NPs and by this maximize the ME coupling effect, a second 500 nm BTO film was deposited subsequently. Finally a 25 nm Au layer was deposited on top of the entire stack to serve as top electrode while the conductive Nb-doped STO substrate is used as bottom electrode.

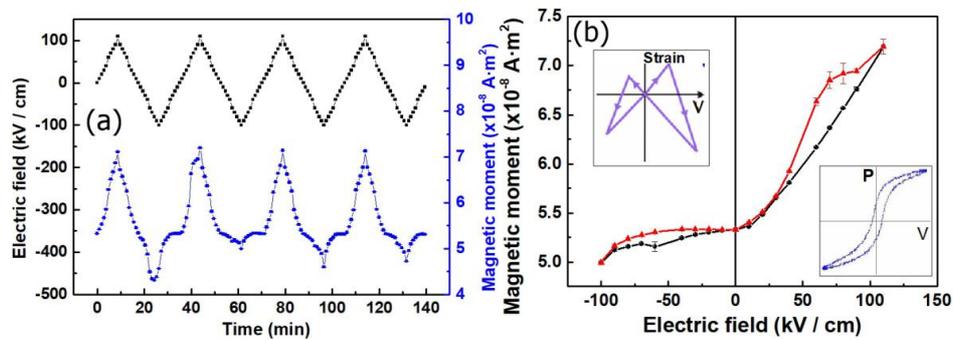

Figure 3. (a) Electric field (black) and magnetic moment (blue) as function of time. (b) Magnetic moment vs. electric field at 300 K and 5 mT. The black curve was measured with increasing DC field from -100 kV/cm to +100 kV/cm, while the red curve was measured in the reversed sequence. The insets show the in-plane strain (upper left) and polarization vs. voltage (bottom right) of the BTO film as adapted from Ref. [21], respectively.

The manipulation of magnetization of the sample was measured as a function of applied electric field. The amplitude of the electric field was varied from -100 kV/cm to +100 kV/cm while a constant magnetic field of 5 mT was set during the measurement. The time dependent electric field and the magnetic moment are displayed in Fig. 3(a). The magnetic moment of the sample changes clearly periodically and synchronized with the electric field which reveals that the electric field directly affects the magnetization of the system. Since both the substrate, the BTO layers and Au layers contribute only with a negligible magnetic moment, the magnetic response shown here originates solely from the iron oxide NPs (see supporting information).

This tuning of the magnetization can be interpreted in terms of an indirect and a direct ME coupling effect.
As indirect coupling, mechanical strain is induced by the application of an electric field via the piezoelectric properties of the BTO film. This changes the magnetization of the iron oxide NPs due to the strain acting on the NPs and the associated change in magnetostrictive energy of the NPs.
The strain as function of electric voltage is expected to show a butterfly shaped curve which is shown in the left-up corner inset in Fig. 3(b) as a schematic [17].
In addition we assume also a direct ME coupling effect resulting from interface charges between the BTO matrix and the iron oxide NPs. I.e. the induced electric polarization in the BTO film generates surface charges at the surface of the BTO film to the NPs. The charges are screened by an equal number of charges of opposite sign at the surface of the NPs. The screening charges would then be spin polarized due to a net FM exchange interaction inside the NPs resulting in an additional magnetic moment. [22]. Such a scenario would show qualitatively a behavior as we observe in the inset at the bottom right in Fig 3(b). The curve magnetization vs. electric fields shows an asymmetric shape which can be decomposed in a part of a ferroelectric hysteresis loop and butterfly shaped loop corresponding to piezoelectricity.

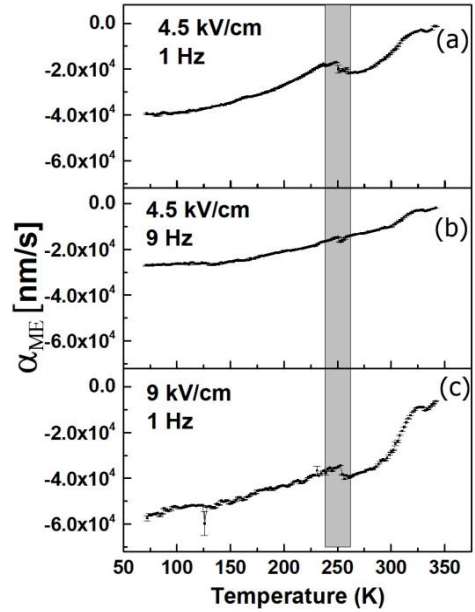

Figure 4. ME coefficient $\alpha_{ME}$ vs. temperature measured at a constant magnetic field of 5 mT and an AC electric field with different frequencies and amplitudes as indicated in the panels. The gray bar marks the phase transition region of the BTO film.

Apart from DC electric field measurements, the effect of an AC electric field onto the magnetic properties of the sample was also probed. The temperature dependent magnetoelectric AC susceptibility (MEACS) coefficient $\alpha_{ME}$ was measured with the AC electric field with a frequency $f = 1$ or 9 Hz and amplitude $U_0 = 4.5$ or 9 kV/cm, respectively. The $\alpha_{ME}$ directly probes the MEC effect at a chosen frequency [17].

$\alpha_{ME}$ as a function of temperature for various AC electric fields is depicted in Figure 4(a-c). The sample was first investigated upon cooling with an AC electric field amplitude of 4.5 kV/cm, with a frequency of 1 Hz and a constant magnetic field of 5 mT (Figure 4(a)). The $\alpha_{ME}$ is negative and decreases smoothly as the temperature decreases. Moreover, a jump at around 250 K is observed. The magnetization jump is very likely due to the MEC effect induced by strain from the BTO. Similarly prepared BTO films show a phase transition at 248 K as found from permittivity vs. temperature measurements [23].

When further increasing the frequency to 9 Hz, both the amplitude of $\alpha_{ME}$ and the signal jump at 250 K become smaller (Figure 4(b)) which indicates that the MEC effect has a frequency dependence and that qualitatively the ME coupling cannot follow the AC field excitation due to an intrinsic speed of the coupling. Correspondingly, the imaginary part of the coefficient increases at 9 Hz compared to 1 Hz which represents an increasing dissipation as the frequency increases (not shown) [23].

It is expected that a stronger electric field of 9 kV/cm would enhance both the amplitude of $\alpha_{ME}$ and the jump height (Figure 4(c)), because a larger electric field will enhance the strain amplitude of BTO film and induce more charges at the interface between BTO film and iron oxide NPs. The absolute value of $\alpha_{ME}$ does increase but the jump height is basically similar as in the case of 4.5 kV/cm. Obviously the action onto the NPs saturates already at 4.5 kV/cm and cannot be enhanced further.

**Conclusions**

We prepared artificial ME heterostructures composed of iron oxide NPs which are embedded in BTO thin films. A MEC effect of this composite is observed as demonstrated by measurements of the magnetization vs. temperature and vs. electric field and using MEACS. A combined strain and interface charge co-mediated mechanism is responsible for the MEC effect. Such NP based ME composites provide an inexpensive way for the fabrication of electronic devices by self-assembly techniques.

**Acknowledgements**

Financial support from China Scholarship Council (CSC) is gratefully acknowledged. We would like to thank Prof. C. Schneider (PGI-6) for providing us the possibility to use their equipment. We thank Prof. R. Waser (PGI-7) for the opportunity to use their SEM and Jochen Friedrich for technical help. We also thank Dr. Markus Schmitz for fruitful discussions and Berthold Schmitz for valuable help.

**Supplementary**

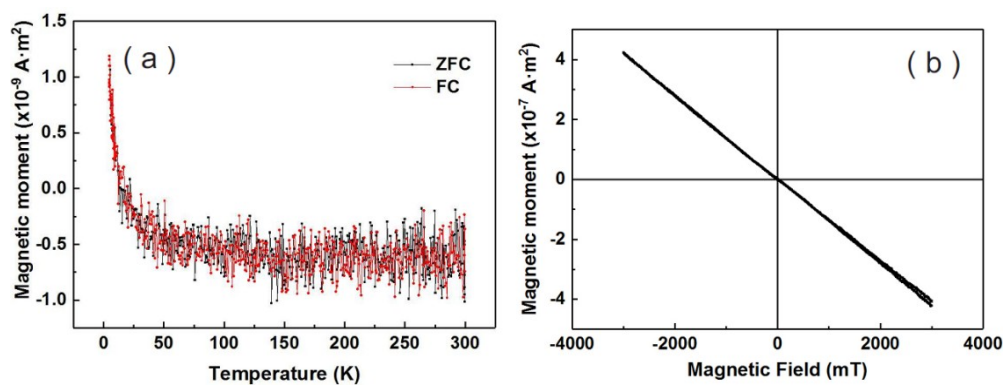

Figure S1. Magnetic moment of the sample BTO film / Nb-doped STO substrate. (a) ZFC and FC curves at 5mT. (b) Magnetic moment vs. magnetic field at 300 K.

The temperature dependent magnetization was measured as ZFC and FC curves and the results are presented in Figure S1(a). One observes only a superposition of diamagnetism at the temperature of interest.

The magnetic field dependent magnetization at room temperature is shown in Figure S1(b). A regular diamagnetic behavior of the BTO film and of the substrate is found as a linear curve of the magnetic moment vs. magnetic field. Moreover, the magnetic moment of the BTO film is in the range of $10^{-9}$ A•$m^2$, which is two orders magnitude smaller than the magnetic moment of iron oxide NP layers.